\documentclass[aps,prl,reprint,superscriptaddress]{revtex4-1}
\usepackage{graphicx}
\usepackage{epstopdf,gensymb}
\usepackage{color}
\usepackage{subfigure}
\usepackage{dcolumn}
\usepackage{bm}
\usepackage{float}

\begin{document}

\title{Generation of `Super-Ponderomotive' Electrons due to a non-Wakefield interaction between a Laser Pulse and a Longitudinal Electric Field }
\author{A.P.L.Robinson}
\email{alex.robinson@stfc.ac.uk}
\affiliation{Central Laser Facility, STFC Rutherford-Appleton Laboratory, Didcot, OX11 0QX, United Kingdom}

\author{A.V.Arefiev}
\affiliation{Institute for Fusion Studies, The University of Texas, Austin,Texas 78712, USA}

\author{D.Neely}
\affiliation{Central Laser Facility, STFC Rutherford-Appleton Laboratory, Didcot, OX11 0QX, United Kingdom}

\date{\today}

\begin{abstract}
It is shown that electrons with momenta exceeding the `free electron' limit of $m_eca_0^2/2$ can be produced when a laser pulse and a longitudinal electric field interact with an electron via a non-wakefield mechanism. The mechanism consists of two stages: the reduction of the electron dephasing rate $\gamma-p_x/m_ec$ by an accelerating region of electric field and electron acceleration by the laser via the Lorentz force. This mechanism can, in principle, produce electrons that have longtudinal momenta that is a significant multiple of $m_eca_0^2/2$.  2D PIC simulations of a relatively simple laser-plasma interaction indicate that the generation of super-ponderomotive electrons is strongly affected by this `anti-dephasing' mechanism.
\end{abstract}
\maketitle

The production of highly energetic electrons in the interaction of ultra-intense laser pulses with plasmas~\cite{abs1} is an essential feature of laser-plasma physics that underpins a wide variety of topics ranging from wakefield acceleration~\cite{mangles_2004} through to Fast Ignition inertial confinement fusion \cite{tabak}.  For certain topics the production of electrons with the highest energies possible is a matter of specific interest.  Examples include laser-driven ion acceleration schemes based on energetic electrons~\cite{exp19} (e.g. Target Normal Sheath Acceleration and those closely related mechanisms), x-ray generation~\cite{kneip_2009}, strong field physics \cite{ridgers_2012}, and positron production in high $Z$ targets~\cite{chen2010}.

The question as to how to reach high energies, and in particular how to exceed what might be termed the `free electron' forward momentum limit of $m_e c a_0^2/2$ \cite{boyd,pukhov1} is therefore one of general interest.  The wakefield scheme is one such route to producing high energies, although does not {\it directly} involve the laser field, and is not very well suited to producing high currents of energetic electrons (c.f. laser interactions with dense plasmas \cite{wilks2,chrisman,habara,yu,gaillard}).  Mechanisms that can breach the `free electron' limit and which directly involve the laser field seem to be more subtle, such as the `Direct Laser Acceleration' scheme that takes place in the ion channel produced by transverse ponderomotive expulsion of electrons by the laser pulse. This was first analyzed by Pukhov et al. \cite{pukhov_dla}, and was more recently re-analyzed by Arefiev et al. \cite{arefiev1}.  In what follows we define the term `super-ponderomotive' electron to mean an electron with forward momentum exceeding  $m_e c a_0^2/2$   

In Ref.~\cite{arefiev1}, a {\it general} mechanism of producing super-ponderomotive electrons was put forward. The interaction of an electron with a laser field in vacuum will have an integral of motion of the form $\gamma-p_x/m_ec = R$ \cite{boyd,pukhov1}.  It is also the case that $\gamma-p_x/m_ec$ is the dephasing rate of the electron, and the `free electron' momentum limit arises from $R=1$ in the absence of any fields apart from the laser field.  Arefiev et al. showed that, in the {\it specific} case of the ion channel, the transverse electrostatic field in the ion channel can reduce the dephasing rate, and thus super-ponderomotive electrons can be produced.  However there is no {\it a priori} reason why only a transverse electric field can reduce $R$ below unity.

In this Letter we show that this very general mechanism extends to the longitudinal field as well, i.e. collective electric field in the direction of laser propagation. Longitudinal electric fields are naturally established in laser-plasma interactions by ponderomotive displacement of electrons, so they are a clear `next candidate' for extra fields that could reduce the dephasing rate. We show that super-ponderomotive energies can result from electron interactions with spikes of relatively weak longitudinal electrostatic field. In contrast with the wakefield acceleration, the axial acceleration in this case is insignificant in terms of the direct energy gain. Instead, the role of the longitudinal field is a reduction of dephasing that subsequently allows for an extended interaction with the laser field and leads to a significant energy enhancement. The strongest increase in energy occurs if the interaction with the longitudinal field is terminated as the electron passes through a zero in the vector potential of the laser field. This suggests similarities with other phenomena where vector potential zeroes are critical \cite{zvp1,tpb2,tpb3}. This mechanism can work in a region of underdense to near-critical plasma formed in front of a dense target. Therefore, it may be the case that the mechanism is partly responsible for the production of highly energetic electrons in current and extant experiments. To show that this mechanism can naturally occur in laser-target interactions, we also present 1D and 2D particle-in-cell (PIC) simulations, which show the anti-dephasing mechanism producing super-ponderomotive electrons in fully self-consistent calculations.

Consider the dynamics of a single electron in an essentially 1D configuration in which it interacts with a plane electromagnetic wave described by the vector potential, ${\bf A} = \left[0,0,A\right] = \left[ 0, 0,A_0\cos\left(\omega_L\tau\right) \right]$,
where $\tau = t - x/c$ and $\omega_L$ is the wave frequency. The electric and magnetic fields are related to the vector potential via ${\bf E} = -\partial_t{\bf{A}}$ and ${\bf B} = \nabla \times {\bf A}$, so the electric field of this wave is polarized in the $z$-direction.  We also consider the case where a longitudinal electric field, $E_x$, is present.  The equations of motion of the electron that need to be considered are:
\begin{eqnarray}
&& \frac{dp_x}{dt} = -eE_x + ev_zB_y, \label{pxeqn} \\
&& \frac{dp_y}{dt} = 0, \label{pyeqn} \\
&& \frac{dp_z}{dt} = -eE_z - ev_xB_y, \label{pzeqn} \\
&& \frac{d\gamma}{dt} = -\frac{ev_zE_z}{m_ec^2} -\frac{ev_xE_x}{m_ec^2}. \label{nrgeqn}
\end{eqnarray}

From the definition of $\tau$, one can differentiate to obtain, $d\tau/dt = 1 - v_x/c$,
and this can then be used to write the field components as $E_z = -\partial_\tau{A}$, $B_y = (1/c)\partial_\tau{A}$. These can then be used to obtain, $p_z = eA$, from Eq.~\ref{pzeqn}, which is one of the key integrals of motion. In the absence of $E_x$, another integral of motion is obtained from Eq.s \ref{pxeqn} and \ref{nrgeqn}, namely $\gamma - p_x/m_ec = 1$ (assuming that the electron is initially at rest).  Using this, one obtains $p_x = e^2A^2/2 m_e c$ in the $E_x = 0$ case (i.e. the `free electron' case).  If, however, $E_x = -E$ (where $E$ is a positive constant over some region) then we instead have,
\begin{equation}
\label{dephase}
\frac{d}{d\tau}\left(\gamma - \frac{p_x}{m_ec}\right) = -\frac{eE}{m_ec},
\end{equation}
and from this we can see that $\gamma -p_x/m_ec < 1$.  We can now re-write Eq. \ref{pxeqn} as,
\begin{eqnarray}
&& \frac{dp_x}{dt} = \frac{1}{R}\frac{e^2A}{m_ec}\frac{dA}{dt}+eE, \label{pxeqn3} \\
&& R = \gamma-\frac{p_x}{m_ec}=1 - \frac{eE}{m_ec}\int{d\tau}. \label{pxeqn4}
\end{eqnarray}
From Eqs.~\ref{pxeqn3} one can see that the effect of the accelerating electric field will not only be direct acceleration of the electron (similar to wakefield acceleration), but it will also be a reduction of the dephasing rate $R$. As a result, the electron will gain additional energy from the laser field above that obtained in the free electron case, i.e. it can produce super-ponderomotive electrons. Equation \ref{pxeqn3} also emphasizes that the `${\bf j}\times{\bf B}$' force is not entirely separated from the longitudinal electric force, as the two are linked through the dephasing rate.

If the electric field has a limited spatial extent, then after passing through this spike one will have $E_x =0$,  but it will still be the case that $\gamma-p_x/m_ec =R$, and Eq.~\ref{pxeqn3} can then be directly integrated to give,
\begin{equation} \label{px}
p_x = p_x^\ast + \frac{m_ec}{2}\frac{a^2-a^{\ast{2}}}{R},
\end{equation}
where we have introduced $a = eA/m_ec$, and where $p_x^\ast$ and $a^\ast$ are the longitudinal momentum and normalized vector potential immediately after the interaction with the spike in the electric field.  It is clear that the largest effect will be obtained if the dephasing rate $R$ is significantly reduced and 
the region of interaction with $E_x$ terminates close to a zero in the vector potential ($a^\ast \ll a_0$).  At $a^\ast = 0$, we have $p_z = 0$  and we immediately find that the reduced dephasing rate in this case is given by 
\begin{equation}\label{R}
R = \gamma - p_x^*/m_e c = \sqrt{\left( p_x^* / m_e c \right)^2 + 1} - p_x^* / m_e c.
\end{equation}
To significantly decrease the dephasing rate ($R \leq 0.5$), the longitudinal momentum following the interaction has to be relativistic. Specifically, assuming that $p_x^*/m_e c \gg 1$, we find directly from Eqs.~(\ref{R}) and (\ref{px}) that
\begin{eqnarray}\label{R2}
& R \approx \left. m_e c \right/ 2 p_x^*, \mbox{     }& \mbox{     }
p_x \approx p_x^\ast \left(1 + a^2 \right).
\end{eqnarray}
Therefore, the axial momentum can be enhanced by as much as a factor of $2 p_x^* / m_e c$ compared to the `free  electron' limit even if the change in the axial momentum during the interaction with the static field is relatively small ($a_0^2 \gg p_x^*/m_e c \gg 1$).

\begin{figure}[tb]
\centering
\includegraphics[width = \columnwidth]{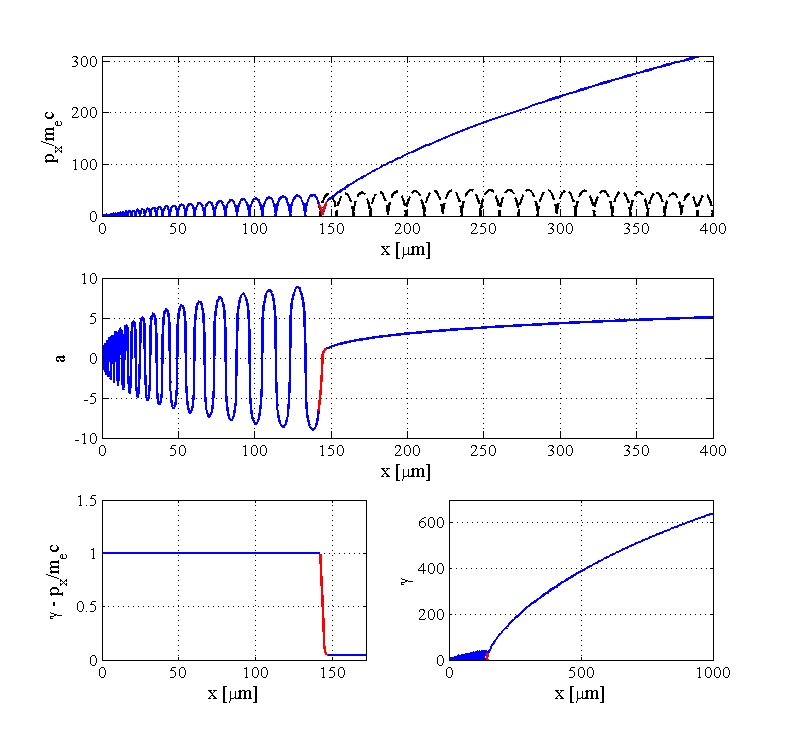} 
\caption{\label{fig:figure1} Electron dynamics in a laser filed [see Eq.~(\ref{laser_1D})] and static field $E_x$ located in the highlighted region. The dashed curve is the axial momentum in the absence of $E_x$.}
\label{fig:figure1}
\end{figure}


To estimate the amplitude of the static field required to reduce the dephasing rate, $R = \gamma-p_x/m_ec$, well below 1, we again use the assumption of a region of uniform electric field. Making use of $d\tau/dt = 1 - v_x/c$ and Eq. \ref{dephase} we have $d\tau/dt = R/\gamma$.  From this the two key equations to consider are Eq.\ref{pxeqn3} and,
\begin{equation}
\label{dephase2}
\frac{dR}{dt} = -\frac{eE}{m_ec}\frac{R}{R+\frac{p_x}{m_ec}}.
\end{equation}
From Eq.\ref{dephase2} we can see that at the zeros in the vector potential ($p_x \approx 0$) we can achieve a rapid fall in $R$, i.e. $dR/dt = -eE/m_ec$.  At the peaks of the laser field ($p_x \approx e^2A^2/2m_ec$) the reduction in dephasing will be much slower, i.e. $dR/dt \approx -(eE/m_ec)(2R/a_0^2)$. One therefore expects, in general, for the largest drops in dephasing to occur around zeroes in the vector potential. To significantly decrease the dephasing rate ($R \leq 0.5$) in a single spike around a zero in the vector potential, one would estimate $\delta{t} \approx 1/(a_0\omega_L)$ which implies $E \approx E_L/2\pi$, where $E_L$ is the amplitude of the lase field. With a more extended electric field, the actual field magnitude required will be significantly less.
                                                                                                     
The insights gained from this analysis can be verified by direct numerical integration of Eq.s \ref{pxeqn}--\ref{pzeqn} along with $dx/dt = p_x/\gamma{m_e}$. Here we take a laser field defined by
\begin{equation} \label{laser_1D}
a_z = a_0\cos(\omega_L\tau)\exp\left[\frac{-(x-ct-x_0)^2}{2c^2t_L^2}\right],
\end{equation}
where $a_0 = 10$, $\lambda = $1$\mu$m, $x_0 =6ct_L$, and $t_L =$40~fs.  The electron is initially at rest at the origin.  A constant longitudinal electric field is equal to $E_x = -0.1 E_L$ at $142 \mu$m$ \ge x \ge 147 \mu$m and it is zero at all other points. The change in the electrostatic potential across this region is roughly 16~MeV, which would result in accelerating the electron to $p_x/m_ec \approx$31 without the laser. The results of a calculation shown in Fig. \ref{fig:figure1} demonstrate that the longitudinal field drives the electron onto a super-ponderomotive trajectory. It is evident from the plot of $a$ at the electron location that the interaction with $E_x$ lasts less than a single oscillation of the laser field (the red segment of the curve). The immediate effect of the axial field is negative, as the axial momentum decreases during the interaction compared to its value calculated for $E_x = 0$. However, the interaction leads to a considerable drop in $\gamma-p_x/m_ec$ and it terminates close to a zero in the vector potential. As a result, a subsequent interaction with the laser field leads to a significant longitudinal acceleration, with the peak momentum in the excess of $10^3 m_e c$. This result is consistent with Eq.~(\ref{px}) since $p_x^*/m_e c \approx 27.5$, $R \approx 0.04$, $a_0 = 10$, and $a^* / a_0 \approx 0.12$.

\begin{figure}[tb]
  \centering
  \subfigure{\includegraphics[scale=0.35]{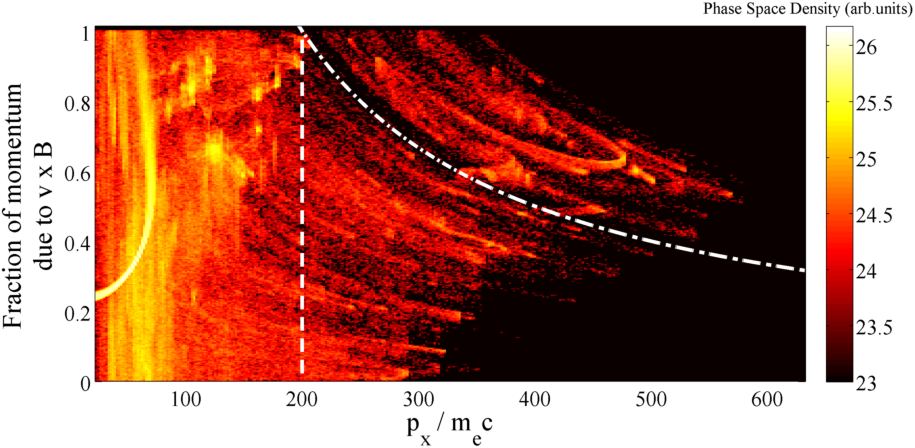}} 
\caption{Electron phase space in 1D PIC simulation at 300~fs (see text) shown as $p_x$ versus the fraction of each macroparticle's momentum due to $ev_zB_y$ acceleration only.  The dashed vertical line indicates the `ponderomotive limit', i.e. $m_eca_0^2/2 $, and the dash-dot line represents the point at which the portion due to $ev_zB_y$ is equal to $m_eca_0^2/2 $.}
\label{Figure_extra}
\end{figure}

In actual laser-plasma interactions, there is a considerable degree of complexity. For example, if the longitudinal electric field does not accelerate the electrons in the direction of the laser pulse, then its effect will be instead to reduce the electron momentum. In order to make a more self-consistent assessment of the importance of this mechanism, we first carried out a parametric study using 1D Particle-In-Cell (PIC) simulations of 100~fs flat-topped laser pulses with $a_0 =$3--20 and $\lambda = 1\mu$m interacting with uniform plasma slabs with densities ranging from 0.01--0.5~$n_c$.  We separately tracked the amount of each macroparticle's axial momentum that was due to $ev_zB_y$ and $-eE_x$. A super-ponderomotive macroparticle with a high fraction of its momentum due to $ev_zB_y$ can only have obtained it from the anti-dephasing mechanism. We observed that, across most of the investigated parameter space, a substantial fraction ($>30\%$) of the electron energy was converted into super-ponderomotive electrons. About 40--60\% of the axial momenta of super-ponderomotive electrons was due to $ev_zB_y$, which shows that the anti-dephasing mechanism is critically important in the generation of these electrons. Figure~\ref{Figure_extra} shows the electron phase space in the form of the momentum fraction due to $ev_zB_y$ against $p_x$ for $a_0$ = 20 and $n_e =$0.1~$n_c$ at 300~fs.  This phase space plot therefore shows both a substantial number of electrons that are super-ponderomotive and that a large fraction of this is due to $ev_zB_y $, hence the anti-dephasing mechanism must be highly significant in these interactions.


To check that the anti-dephasing mechanism could also be observed even when multi-dimensional effects are included, we carried out 2D PIC simulations using the Plasma Simulation Code (PSC). A laser pulse was normally incident significantly underdense hydrogenic plasma with density $n_0 = 8\times$10$^{25}$m$^{-3}$. The length and width of the slab being 200$\mu$m and 160$\mu$m respectively, and the simulation domain with open boundaries was 300$\mu$m by 200$\mu$m (12000 by 2000 cells). The laser pulse had $a_0 =10$, $\lambda = 1 \mu$m, a FWHM width of 8$\mu$m, and a pulse duration of 500~fs.  Denoting the two axes of the simulation domain as $x$ (laser direction) and $y$, the electric field of the laser was polarized in the $z$ direction.

\begin{figure}[tb]
  \centering
  \subfigure{\includegraphics[scale=0.55]{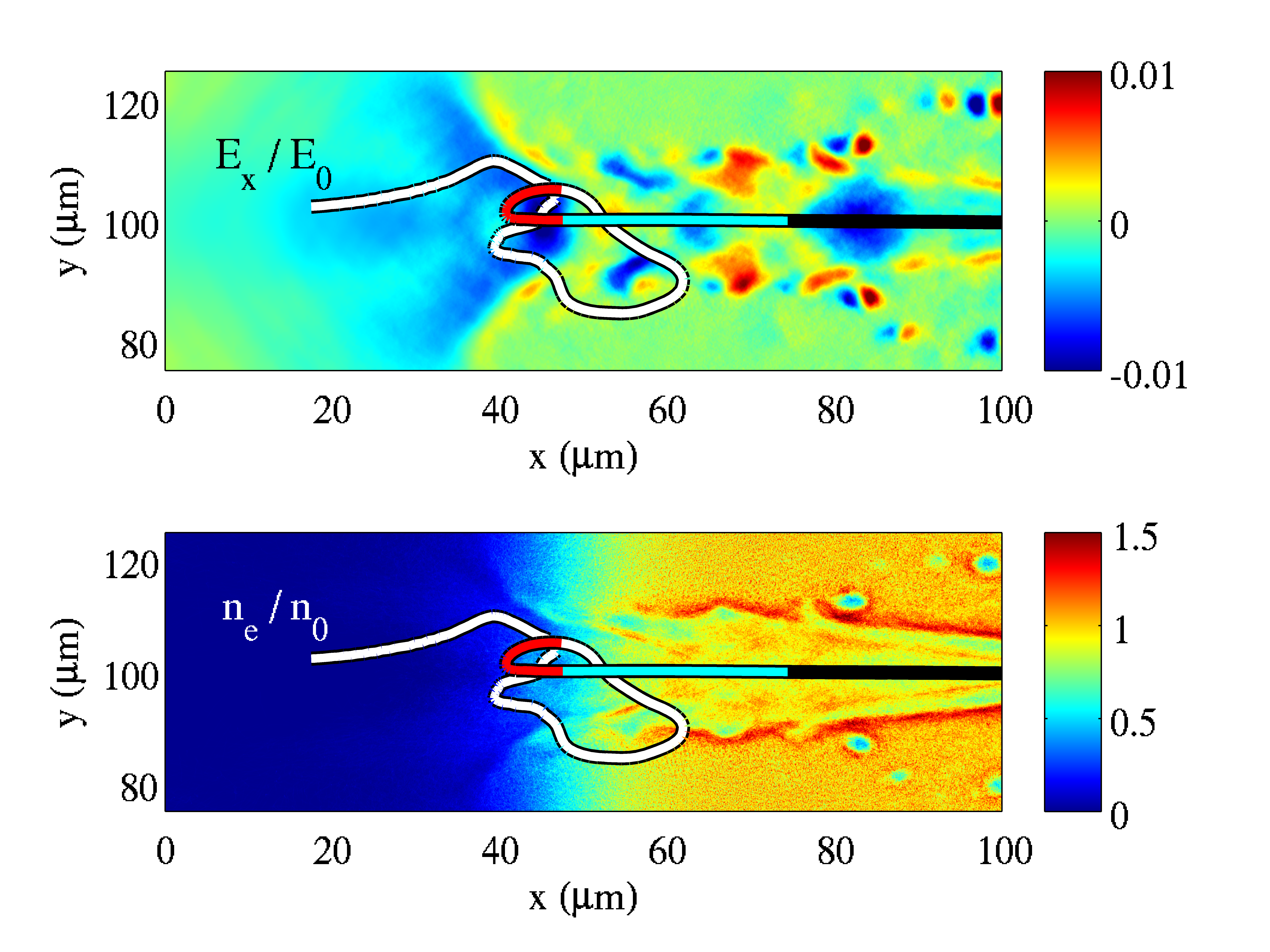}} \\
  \subfigure{\includegraphics[scale=0.50]{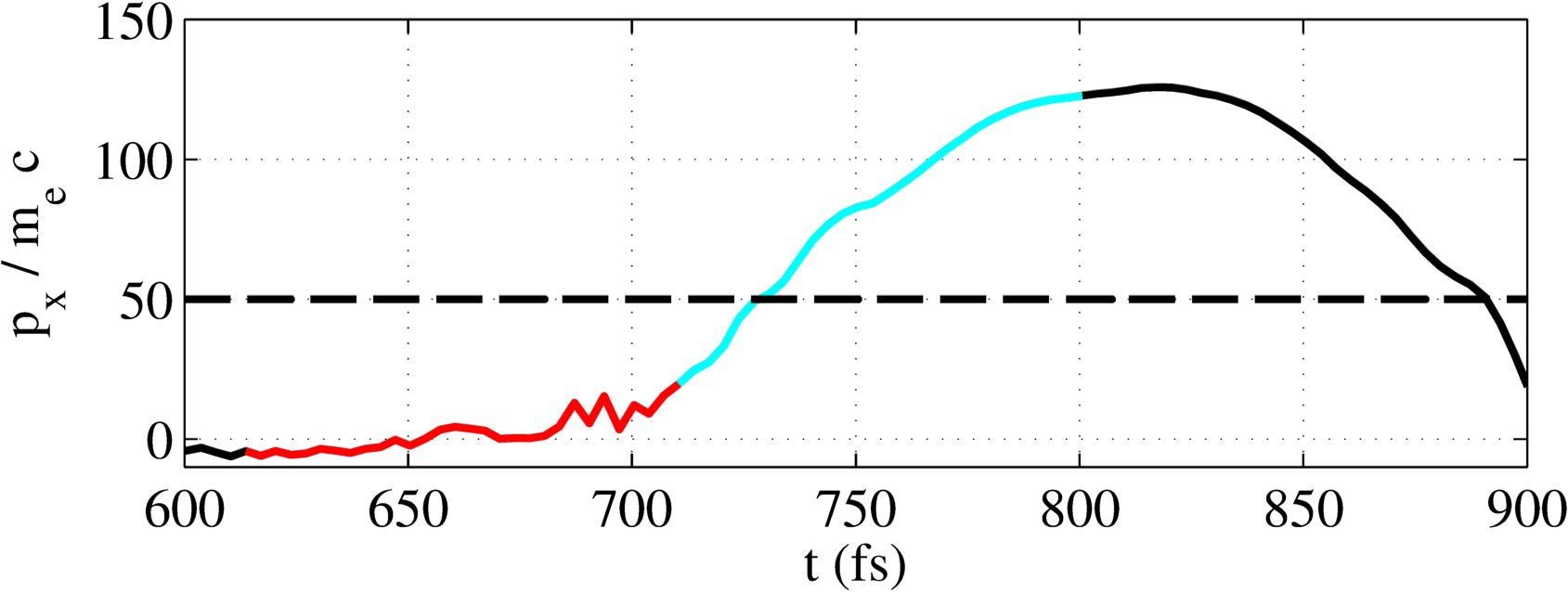}} \\
  \subfigure{\includegraphics[scale=0.50]{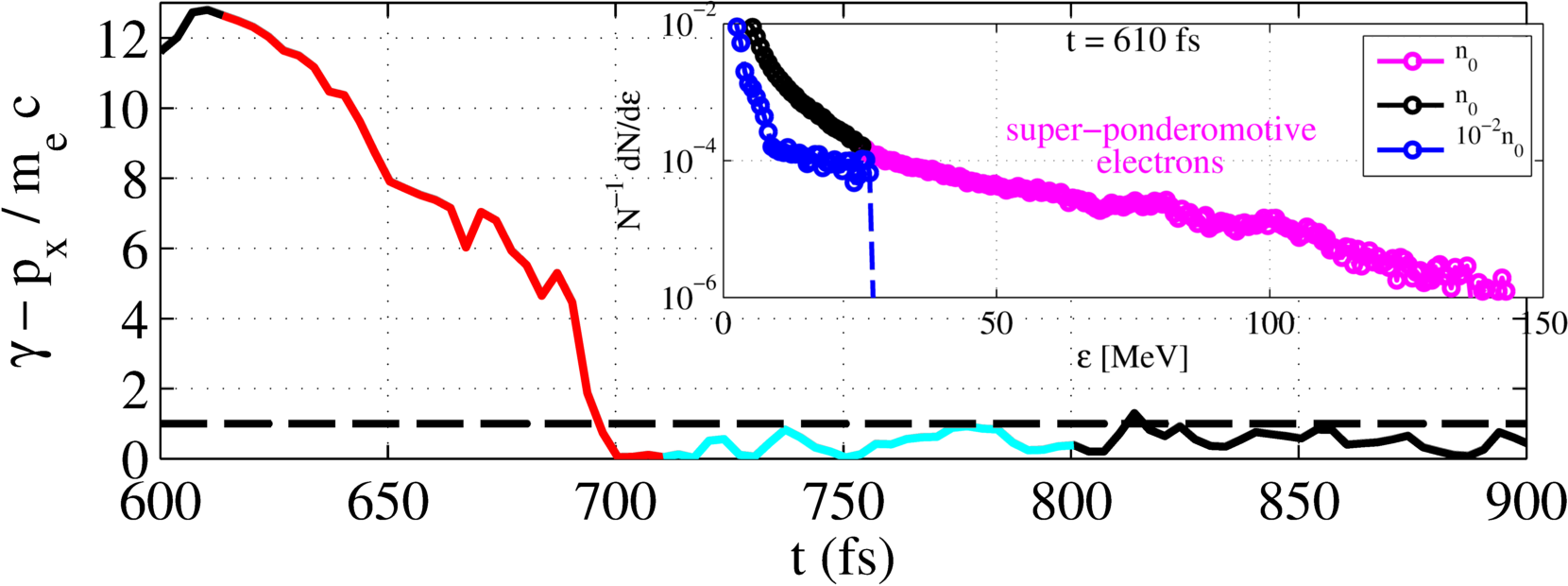}}
\caption{Snapshots of time-averaged axial electric field, electron density, and time evolution of the electron axial momentum and the dephasing from a 2D PIC simulation.}
\label{Figure_2D}
\end{figure}

This pulse is significantly longer than the characteristic time of electron response in the plasma, so that the laser can creates a quasistatic channel shown in Fig.~\ref{Figure_2D}. The channel has a coaxial structure with a positively charged center and negatively charged walls. During the formation of the channel, the electrons from inside the channel are expelled radially by ponderomotive pressure, which causes charge separation. The resulting radial electric field counteracts the expelling force, allowing the electrons to remain in an equilibrium bunched on the periphery of the laser beam. Such channels and the corresponding transverse electric field are routinely observed in simulations of laser interactions with underdense plasmas~\cite{sentoku2006,li2008,sarri2010,friou2012}. However, the fact that such a coaxial structure also produces a quasistatic axial electric field at the channel opening (see Fig.~\ref{Figure_2D}) has been underappreciated. 

In order to examine the role played by the axial field, we have performed a search on the electron data for super-ponderomotive electrons for which $\gamma-p_x/m_ec <$0.05 and $x <$50$\mu$m at least at some point during the electron trajectory. Figure~\ref{Figure_2D} shows a trajectory, axial momentum, and dephasing for just one such electron. At the channel opening, there is a region with a strong quasi-static negative $E_x$. The electron interaction with this field (shown on all plots with a red segment) launches the electron onto a super-ponderomotive trajectory (the subsequent acceleration is shown with a light-blue curve). There is virtually no self-focusing of the laser in this region ($a_0 \approx $10), so the free-electron limit for $\gamma$ is 50. The electron however achieves a peak $\gamma$ exceeding the free-electron limit by a factor of three. The acceleration is preceded by a massive drop in the dephasing rate that occurs during the interaction with $E_x$ when the electron moving against the laser beam is turned around and pushed forward. The time evolution of the axial momentum further emphasizes that this is a two-stage process, since no significant axial acceleration occurs directly during the interaction with the axial field.

The inset in Fig.~\ref{Figure_2D} shows snapshots of electron spectra at densities $n_0$ and $10^{-2} n_0$ normalized to the total number of electrons in the slab. There are copious super-ponderomotive electrons at higher density (they account for 14\% of all electron energy), whereas there are no such electrons at lower density. The total energy absorbed by the electrons has increased by a factor of $1.8 \times 10^4$, while the number of electrons in the slab increased only by a factor of 100. At lower density, the channel is fully evacuated and no electrons sample the axial field at the channel opening. At higher density, new electrons are continuously injected into the channel near the opening and pass through the region of the strong axial field. The continuous injection together with the anti-dephasing mechanism leads to the significant increase of electron heating. 


The electric field of the laser is polarized out of the plane of the simulation, which eliminates the betatron resonance~\cite{pukhov_dla} as a possible explanation for the observed energy gain. We also observe no amplification of the transverse oscillations across the channel, which indicates that the observed effect is not related to the parametric amplification~\cite{arefiev1}. Figure~\ref{Figure_2D} clearly shows that this event is quite prompt, so the underlying mechanism must be able to produce the observed behavior without any gradual build-up. The reduction of dephasing by acceleration in the longitudinal electric field clearly satisfies this key criterion. Later on, we observe a decline in the electron momentum, which illustrates the concern, stated earlier, that the collective fields can also act to increase the dephasing rate.  Note that in this simulation there is no dense foil which will interrupt the acceleration process. We have therefore made no attempt to determine the optimal conditions for exploiting this anti-dephasing mechanism, as these conditions will be highly dependent on the specifics of the laser and target parameters.

In this Letter we have shown that, according to the single electron equations of motion, electrons can be accelerated to momenta in excess of $m_eca_0^2/2$ by an `anti-dephasing' mechanism in which a brief acceleration by a longitudinal electric field that is simultaneously present with the laser pulse reduces $\gamma-p_x/m_ec$ significantly below unity.  The existence and importance of this mechanism was verified by direct numerical integration of the equations of motion and then the effect was confirmed in a fully self-consistent simulation of laser-plasma interaction (laser pulse propagating in under-dense plasma). This mechanism is complimentary to the mechanism of  the parametric amplification of betatron oscillations, so that the combination of the two can produce super-ponderomotive electrons with energies exceeding what is predicted here and in Ref.~\cite{arefiev1}. This work also shows that one cannot simply split electron motion into independent 'wakefield-like' and  'free-electron-like' components, as the anti-dephasing mechanism is due to a subtle interaction between the two.

\begin{acknowledgements}
APLR is grateful for computing resources provided by STFC's e-Science facility. AVA acknowledges the Texas Advanced Computing Center at The University of Texas at Austin for providing HPC resources. AVA was supported by National Nuclear Security Administration Contract No. DE-FC52-08NA28512, U.S. Department of Energy Contract No. DE-FG02-04ER54742, and Sandia National Laboratory Contract No. PO 990947.

\end{acknowledgements}

\end{document}